\documentclass[12pt]{iopart}
\usepackage{graphicx,multirow}

\newcommand{\dint}{\int\!\!\int}

\newcommand{\dr}{\,d{\bf r}}

\newcommand{\hal}{\frac{1}{2}}
\newcommand{\RI}{{\bf R}_I}
\newcommand{\RJ}{{\bf R}_J}
\newcommand{\Ru}{{\bf R}_u}
\newcommand{\rI}{{\bf r}_I}
\newcommand{\rJ}{{\bf r}_J}
\newcommand{\ru}{{\bf r}_u}
\newcommand{\kb}{{\bf k}}
\begin{document}

\title{
Efficient evaluation of the Fourier Transform over products of
Slater-type orbitals on different centers}
\author{T A Niehaus$^{1,2}$, R L\'opez$^{3}$ and J F Rico$^{3}$}
\address{$^{1}$ Bremen Center for Computational Materials Science, 
          University of Bremen, 
          D-28359 Bremen, Germany }
\address{$^{2}$ German Cancer Research Center,
          Dept. Molecular Biophysics,
          D-69120 Heidelberg, Germany } 

\address{$^{3}$ Departamento de Qu\'imica F\'isica Aplicada, Facultat de
  Ciencas C-XIV, Universidad Aut\'onoma de Madrid, 28049 Madrid, Spain}

\date{\today}
\pacs{71.15.-m, 71.15.Ap}
\begin{abstract}
Using the shift-operator technique, a compact formula for the Fourier
transform of a product of two Slater-type orbitals located on
different atomic centers is derived. The result is valid for arbitrary
quantum numbers and was found to be numerically stable for a wide
range of geometrical parameters and momenta. Details of the
implementation are presented together with benchmark data for
representative integrals. We also discuss the assets and drawbacks of
alternative algorithms available and analyze the numerical efficiency
of the new scheme. 
\end{abstract}

\maketitle

\section{Introduction}
The electron-electron interaction as quantified in Coulomb or exchange
integrals is at the heart of every quantum mechanical treatment of
condensed matter. Due to the simple structure of the Coulomb
operator in reciprocal space, Fourier transform techniques allow for
the transformation of the double integral over real space into a
compact single integral in momentum space. Let us consider a typical
two-electron repulsion integral as an example
\begin{equation}
  \label{ft1}
  I = \dint \phi_\mu ({\bf r}-{\bf R}_A) \phi_\nu ({\bf r}-{\bf R}_B)\,
  \frac{1}{|{\bf r} -{\bf r}'|}\, \phi_\alpha ({\bf r}'-{\bf R}_C)
  \phi_\beta ({\bf r}'-{\bf R}_D)  \dr \dr'.
\end{equation}
If we denote the Fourier transform of the product of orbitals $
\phi_\mu({\bf r}-{\bf R}_A)$ and $ \phi_\nu({\bf r}-{\bf R}_B)$ by
$\phi_{\mu\nu}({\bf R}_A,{\bf R}_B,{\bf q})$, we may very schematically
write (thorough definitions follow later)
\begin{equation}
  \label{ft2}
  I \propto \int \,\phi_{\mu\nu}({\bf R}_A,{\bf R}_B,-{\bf q})\,
  \frac{1}{q^2}\, \phi_{\alpha\beta}({\bf R}_C,{\bf R}_D,{\bf q})
  \,d{\bf q}.
\end{equation}
Methods along these lines date back
at least to Bonham et al.\ \cite{bon64,bon65} and are the de-facto standard
for systems with translational symmetry. For these boundary
conditions, plane waves are the most natural type of basis functions,
thanks also to their trivial behavior under Fourier transformation.

In recent years, correlated electronic structure methods like the GW
approximation of Hedin \cite{hed65}, which were originally developed
in the context of band structure calculations, are now also applied to
systems where translational symmetry is broken. Examples of this
kind include super lattices, defects, surfaces and even atoms or
molecules \cite{roh98, roh00, oht01, nie05, nea06}. Obviously, atomic
orbital basis sets are more 
appropriate in these situations and reduce the number of basis
functions needed to achieve a certain accuracy considerably. Among the
localized basis sets, Slater and Gaussian-type orbitals are the most
prominent. The latter have the advantage that the Fourier transform of
orbital products is relatively easy to obtain, while usually much less
Slater than Gaussian-type orbitals are required to represent
atomic or molecular electron densities. In the context of the Fourier
transform methods mentioned above, the choice of Slater versus
Gaussian is hence intimately connected with the ability to perform the
Fourier transform of basis function products efficiently.

A direct numerical quadrature of the three dimensional integral over
reciprocal space using Fast Fourier Transform techniques is not
advisable due to high memory consumption, low computational speed and
very limited numerical accuracy. As an alternative, Slater functions
may be fitted to a fixed linear combination of Gaussian-type orbitals like
done for example in the popular Pople basis sets often employed in
quantum chemistry \cite{pop70, pis80}. However, in this case the electronic
structure calculation could have been performed entirely in terms of
Gaussians in the first place with additional variational
freedom. Several attempts to directly perform the intricate integration of Slater
products analytically are documented in the literature. While the
mentioned original work of Bonham \cite{bon64,bon65} was restricted to s-type functions, Bentley
and Stewart \cite{ben73} derived an expression for arbitrary angular
momentum states involving an infinite series. Later, Junker
\cite{jun80} obtained a result in terms of finite sums and
one-dimensional numerical integrations, while Straton
\cite{stra87,stra88} was able to provide general formulas for the Fourier
transform of the product of more than two  orbitals.

This earlier work meets
some but not all desired properties of a general solution. The latter
include the validity for arbitrary quantum numbers, the possibility
for a straightforward implementation on a computer as well as
high numerical efficiency and stability. Moreover, the solution should
be amenable to partial wave analysis, in order to allow for an
efficient evaluation of two-electron integrals in a second step.
 This point is maybe not so important for periodic systems, where
quadrature of the remaining integral over reciprocal space in
(\ref{ft2}) may be
accomplished by summation over {\em special k-point} meshes
\cite{mon76}, involving only a small number of integrand evaluations. For
finite systems however, this point becomes crucial. Here it should be
mentioned that the efficient evaluation of Fourier
transforms is only a first step in a fast computation of multi-centre
integrals.

An approach which combines most of the above mentioned merits was
proposed by Trivedi and Steinborn \cite{triv83}. The authors
provide formulas for the Fourier transform over products of so-called 
B-functions. These B-functions can be transformed into Slater-type
orbitals without loss of generality and accuracy. Subsequently, this
technique was used by Grotendorst and Steinborn \cite{gro88} to evaluate a variety
of multi-centre integrals required in electronic structure
calculations.  Alternative
representations of the transforms are given in \cite{gro85}
and in the dissertation of Homeier \cite{hoh90}, which also contains a deeper
discussion of the B-function formalism together with numerical results
and benchmark data.

In this work, we
propose an alternative to the Trivedi-Steinborn formula, which is 
{\em directly} formulated in terms of Slater-type orbitals. The
derivation is based on the shift-operator technique, which is discussed
in the next section. The approach may be seen as
a generalization of a recent result for overlap integrals \cite{ema08}
and meets all important criteria established above.

The more general aim of this contribution is to facilitate the utilization of
Slater-type orbitals in the simulation of periodic and quasi-periodic
systems. Currently, only a very limited number of codes employs this
kind of basis set \cite{vel01, fer89}, due to the apparent
difficulties in the numerical implementation. The
development of adapted algorithms is therefore of key importance in order to unveil the
well known benefits of Slater-type orbitals in atomistic calculations.

\section{Definitions and the shift-operator approach}
We consider real unnormalized Slater-type orbitals (STO) of the form:
\begin{equation}
  \label{sto}
\chi_{nlm}({\bf r}, \zeta) = r^{n-1} e^{-\zeta r}  z^m_l({\bf r}),
\end{equation}
where $n = \tilde{n} -l$ in terms of the principle quantum number
$\tilde{n}$. The regular harmonics $z^m_l$ are related to the more
familiar real spherical harmonics $\tilde{Y}_{lm}$: 
\begin{eqnarray}
  \label{zlm}
z^m_l({\bf r}) &=& r^l \tilde{Y}_{lm}({\bf \hat{r}})\\
 \tilde{Y}_{lm}({\bf \hat{r}}) &=& (-1)^m P^m_l(\cos\theta)\cos(m \phi)
      \quad m \ge 0 \nonumber\\\nonumber
 \tilde{Y}_{l-m}({\bf \hat{r}}) &=& (-1)^m P^m_l(\cos\theta)\sin(m \phi)
      \quad m < 0,
\end{eqnarray} 
where $P^m_l(\cos\theta)$ denote associated Legendre polynomials as
defined in \cite{gra80}.
Square normalized STO $\bar{\chi}_{nlm}$ are readily obtained
as:
\begin{equation}
  \label{norm}
  \bar{\chi}_{nlm} = \sqrt{\frac{ (2l + 1)}{2\pi (1+\delta_{m0})}
    \frac{(l-|m|)!}{ (l+|m|)!} \frac{ (2\zeta)^{2n+2l+1}}
{ (2n+2l)!}} \,\chi_{nlm}. 
\end{equation}

The idea of the shift-operator approach is to
evaluate the desired integral of interest, e.g., overlap or  two-electron
repulsion, first for the simplest STO of s-symmetry, for which an
quadrature is often relatively easy. In a second step, the quantum
numbers are then raised by operators that involve derivatives with
respect to parameters of the integral, like decay constants and
inter-center distance. The benefit of such raising and lowering operators
in the solution of molecular integrals was recognized quite early and
exploited by various authors \cite{bar51,gui79,fie80,wen83,fer00,fer01}. 

In this approach, a STO centered at 
$\RI$ as a function of $\rI = {\bf r} - \RI$ may be written as:
\begin{equation}
  \label{soa}
  \chi_{nlm}(\rI, \zeta) = \Omega^n_{lm}(\nabla_I)  \frac{e^{-\zeta
      r_I}}{r_I}, \end{equation}
with $\nabla_I$ denoting the vector $(\partial / \partial
X_I, \partial / \partial Y_I,  \partial / \partial Z_I)$ and 

\begin{equation}
\label{znab}
\Omega^n_{lm}(\nabla_I) =  z^m_l(\nabla_I)
\left(-\frac{\partial}{\partial \zeta}\right)^n 
 \left(-\frac{1}{\zeta} \frac{\partial}{\partial \zeta}\right)^l. 
\end{equation}
A detailed discussion of the properties of $z^m_l(\nabla_I)$ and
related differential operators is provided in a recent review by
Weniger \cite{wen05}.

The form in (\ref{znab}) is now used to construct the Fourier transform of two-center
STO products which are in the focus of this work:
\begin{eqnarray}
  \label{ft}
\fl
  I^{n_1 l_1 m_1}_{n_2 l_2 m_2}(\kb,\zeta_1,\zeta_2,\RI,\RJ) &=& \int
  \dr e^{i \kb {\bf r}} \chi_{n_1 l_1 m_1}(\rI, \zeta_1) \chi_{n_2 l_2
    m_2}(\rJ, \zeta_2) \\
&=&\Omega^{n_1}_{l_1 m_1}(\nabla_I)\Omega^{n_2}_{l_2 m_2}(\nabla_J)  \int
  \dr e^{i \kb {\bf r}} \chi_{000}(\rI, \zeta_1) \chi_{000}(\rJ, \zeta_2).
\end{eqnarray}

The shift-operator approach is applicable if the basic integrals
($I^{000}_{000}$ in our case) have a closed form which can be easily
differentiated with respect to the outer parameters. The next section
shows that this is indeed the case for the present Fourier transform.

\section{The basic integral }
As shown by Rico and co-workers \cite{ric07}, the product of two
s-type STO can be expressed as an one-dimensional integral which is
suitable for further manipulations:
\begin{equation}
  \label{yyy}
\fl
 \chi_{000}(\rI, \zeta_1) \chi_{000}(\rJ, \zeta_2) = \frac{1}{\pi}\int_0^1 du \left[ u
   (1-u) \right]^{-\frac{3}{2}} \zeta_u^2 \,
\hat{k}_{-1}\left(\zeta_u \sqrt{R^2 + \frac{r_u^2}{u(1-u)}} \right),\quad   
\end{equation}
with ${\bf R} = \RJ -\RI,\, \Ru = u \RJ  + (1-u) \RI,\, \ru = {\bf r}
- \Ru,\, \zeta_u^2 = \zeta_1^2 u + \zeta_2^2 (1-u),$ 
and $\hat{k}_{\nu}(x) = x^\nu K_\nu(x)$, where $K_\nu(x)$ is the
modified Bessel function of the second kind, often also  called
McDonald function. 

After insertion of (\ref{yyy}) into (\ref{ft}) and change of
the integration variable to ${\bf r}_u$, the angular integration is
readily performed by expanding the exponential in partial waves
\begin{equation}
  \label{parw}
  e^{i{\bf k} {\bf r}} = \sqrt{\frac{\pi}{2kr}} \sum_{l=0}^\infty i^l
  (2l+1) P_l(\cos \theta) J_{l+ \frac{1}{2}} (k r),\quad  \theta =
  \angle \left({\bf k},{\bf r}\right),
\end{equation}
and using the fact that the remainder of the integrand has
s-symmetry. The result reads:
\begin{eqnarray}
  \label{int1}
\fl
  I^{000}_{000}(\kb,\zeta_1,\zeta_2,\RI,\RJ) = \frac{4}{k} \int_0^1
  du\, e^{i \kb \Ru} \left[ u
   (1-u) \right]^{-\frac{3}{2}} \zeta_u^2 \\\times  \int_0^\infty dr_u r_u \sin(kr_u) \hat{k}_{-1}\left(\zeta_u \sqrt{R^2 + \frac{r_u^2}{u(1-u)}} \right).
\end{eqnarray}
The remaining radial integral is known \cite{gra80eq}, which leads to
the final result for the basic integral:
\begin{equation}
  \label{basint}
\fl
   I^{000}_{000}(\kb,\zeta_1,\zeta_2,\RI,\RJ) = \sqrt{8 \pi} R \int_0^1
  du\, e^{i \kb \Ru} \hat{k}_{-\frac{1}{2}}\left(R \sqrt{\zeta_u^2 + k^2
    u (1-u)} \right).\quad
\end{equation}
For vanishing momentum transfer this formula reduces to the known
result for the corresponding  overlap integral as given, e.g., by Ema et al.\
\cite{ema08}. For this special case, the pending integral may be
solved analytically and is related to confluent hypergeometric
functions. In general however, an evaluation based on numerical
integration is unavoidable at this point.

\section{Transforms for higher quantum numbers}
\label{trahigh}
Using the shift-operator approach, Fourier transforms for higher
quantum numbers may now be written as:
\begin{equation}
  \label{high}
\fl
  I^{n_1 l_1 m_1}_{n_2 l_2 m_2}(\kb,\zeta_1,\zeta_2,\RI,\RJ) =
  z^{m_1}_{l_1}(\nabla_I) z^{m_2}_{l_2}(\nabla_J) \int_0^1   du e^{i
    \kb \Ru} h^{n_1 l_1}_{n_2 l_2}(k,\zeta_1,\zeta_2,R,u)
\end{equation}
with
\begin{eqnarray}
\label{hterm}
\fl
 h^{n_1 l_1}_{n_2 l_2}(k,\zeta_1,\zeta_2,R,u) = 
\sqrt{8 \pi} R  \left(-\frac{\partial}{\partial \zeta_1}\right)^{n_1} 
 \left(-\frac{1}{\zeta_1} \frac{\partial}{\partial
     \zeta_1}\right)^{l_1} \times \nonumber\\ 
\left(-\frac{\partial}{\partial \zeta_2}\right)^{n_2} 
 \left(-\frac{1}{\zeta_2} \frac{\partial}{\partial
     \zeta_2}\right)^{l_2}  \hat{k}_{-\frac{1}{2}}\left(R \sqrt{\zeta_u^2 + k^2
    u (1-u)} \right),
\end{eqnarray}
where, as shown in the \ref{further}, the derivatives with respect to the
decay constants are relatively easy to perform. The action of the
solid harmonics on the integral is more involved and requires further
consideration. We proceed by introducing the equality
\begin{equation}
  \label{dzlm}
  z^{m}_{l}(\nabla) (f\, g) = \sum_{l'=0}^l 
  \sum_{m'=-(l-l')}^{l-l'} 
  \sum_{m''=-l'}^{l'} d^{l m}_{l' m' m''} \left(z^{m'}_{l-l'}(\nabla)\, f \right) 
  \left(z^{m''}_{l'}(\nabla)\, g \right)    
\end{equation}
for arbitrary functions $f({\bf R}), g({\bf R})$. This relation is proven
in the \ref{leib} using the Leibniz rule for the differentiation of products
together with the completeness and orthogonality relations of
spherical harmonics. Alternative proofs were given by Dunlap
\cite{dun90} and Weniger \cite{wen05}. 

Values for the coefficients $d^{l m}_{l' m' m''}$
can be obtained by straightforward differentiation for small quantum
numbers. In general, the use of symbolic computation software allows
the determination once and for all. Special cases include $d^{lm}_{0 m'
  m''} = \delta_{m' m}  \delta_{m'' 0}$ and $d^{lm}_{l m'
  m''} = \delta_{m' 0}  \delta_{m'' m}$.

Applying (\ref{dzlm}) to (\ref{high}), we arrive at 
\begin{eqnarray}
\fl
  I^{n_1 l_1 m_1}_{n_2 l_2 m_2}(\kb,\zeta_1,\zeta_2,\RI,\RJ) =  
 \sum_{{l'_1}=0}^{l_1}  i^{l_1-{l'_1}}
  \sum_{{m_1}'=-({l_1}-{l'_1})}^{{l_1}-{l'_1}} 
  \sum_{{m_1}''=-{l'_1}}^{{l'_1}} d^{{l_1} {m_1}}_{{l'_1} {m_1}'
    {m_1}''} \times \nonumber \\\fl
  \sum_{{l'_2}=0}^{l_2}  i^{l_2-{l'_2}}
  \sum_{{m_2}'=-({l_2}-{l'_2})}^{{l_2}-{l'_2}} 
  \sum_{{m_2}''=-{l'_2}}^{{l'_2}} d^{{l_2} {m_2}}_{{l'_2} {m_2}'
    {m_2}''}
 \int_0^1 du \left[ u^{l_1-{l'_1}}
   (1-u)^{l_2-{l'_2}} 
z^{{m_1}'}_{l_1-{l'_1}}({\bf k}) 
z^{{m_2}'}_{l_2-{l'_2}}({\bf k}) 
  e^{i \kb \Ru} \right] \times\nonumber \\
\left[ z^{{m_1}''}_{{l'_1}}(\nabla_I)
  z^{{m_2}''}_{{l'_2}}({\boldmath \nabla}_J) h^{n_1 l_1}_{n_2
    l_2}(k,\zeta_1,\zeta_2,R,u) \right],  \label{high2}
\end{eqnarray}
where we used the homogeneity of regular harmonics and the fact
that plane waves are eigenfunctions of the momentum operator.
The remaining derivation parallels the work of Ema et al.\ \cite{ema08}
on overlap integrals and we will follow the nomenclature used there as close as
possible to facilitate comparison.
 Since $ h^{n_1 l_1}_{n_2
    l_2}$ in
the last line of (\ref{high2}) depends only on the norm of ${\bf
  R}$, the following theorem may be applied which goes back to Hobson \cite{hob65} 
\begin{eqnarray}
  \label{hobson}
\fl
 z^{{m_1}}_{{l_1}}(\nabla_I)
  z^{{m_2}}_{{l_2}}({\boldmath \nabla}_J) f(R) &= (-1)^{l_1}
  \sum_{k=0}^{L_{<}} \frac{2^{-k}}{k!} \left[ \nabla^{2k}
    z^{{m_1}}_{{l_1}}({\bf R})
  z^{{m_2}}_{{l_2}}({\bf R}) \right]
\left(\frac{1}{R} \frac{\partial}{\partial R}\right)^{l_1 +l_2 -k}
f(R)\\
\label{hobson2}
&= (-1)^{l_1}
  \sum_{k=0}^{L_{<}} {\cal P}_k^{l_1 m_1 l_2 m_2} ({\bf R})
\left(\frac{1}{R} \frac{\partial}{\partial R}\right)^{l_1 +l_2 -k}
f(R)
\end{eqnarray}
Here $L_{<} = {\rm min}(l_1,l_2)$
and the ${\cal P}_k^{l_1 m_1 l_2 m_2}$ are given by
\begin{eqnarray}
  \label{pl}
\fl
  {\cal P}_k^{l_1 m_1 l_2 m_2} ({\bf R}) &=& \frac{2^k}{k!}
  \sum_{l=k}^{L_<} \frac{l! \Gamma(l_1 + l_2 -l + 3/2)
    R^{2l-2k}}{(l-k)! \Gamma(l_1 + l_2 -l -k + 3/2)}
\sum_m c^{l_1 m_1 l_2 m_2}_{l_1 + l_2 -2l\, m} z^{m}_{l_1 +
  l_2 -2l}  ({\bf R}), 
\end{eqnarray}
where the coefficients $c^{l_1 m_1 l_2 m_2}_{l_1 + l_2 -2l\, m}$ are
directly related to real Gaunt coefficients (For a detailed derivation
of (\ref{hobson}) to (\ref{pl}) see \ref{hobs}). 

Next we define the quantity $\tilde{{\cal S}}^{n_1 l_1 n_2
  l_2}_{{l'_1} {l'_2} k}$ (this is a generalization of ${\cal S}^{n_1
  l_1 n_2 l_2}_k$ in the work of Ema et al.\ \cite{ema08}), which is
further discussed in the \ref{further}:
\begin{eqnarray}
  \label{snl}
\fl
 &\tilde{{\cal S}}^{n_1 l_1 n_2
  l_2}_{{l'_1} {l'_2} k}({\bf k},\zeta_1,\zeta_2,\RI,\RJ)
\nonumber
\\ &=   \int_0^1 du\,   e^{i \kb \Ru}
u^{l_2-l'_2} (1-u)^{l_1-l'_1} \left[ \left(-\frac{1}{R}
    \frac{\partial}{\partial R}\right)^{l'_1 +l'_2 -k}  h^{n_1 l_1}_{n_2
    l_2}(k,\zeta_1,\zeta_2,R,u) \right]\nonumber\\
&= (-1)^{l'_1 + l'_2 -k} R^{1+2(l_1+l_2 -l'_1 -l'_2)} \sum_{i=\lfloor\frac{n_1+1}{2}\rfloor}^{n_1}
\sum_{j=\lfloor\frac{n_2+1}{2}\rfloor}^{n_2}  c^{n_1}_ i(\zeta_1) c^{n_2}_j(\zeta_2)   \nonumber\\
&\quad \times \sqrt{8\pi}\int_0^1 du\,   e^{i \kb \Ru}
u^{\mu} (1-u)^{\nu}
\hat{k}_{\alpha}\left( R \sqrt{\zeta_u^2 + k^2
    u (1-u)}  \right),
\end{eqnarray}
with $\lfloor r \rfloor$ denoting the integer part of $r$ and 
\begin{eqnarray}
\label{numu}
\fl
  \mu = l_1+ l_2-l'_2 +i ; \quad \nu = l_1+l_2-l'_1 +j ; \quad \alpha=
  -1/2-l_1-l_2+l'_1+l'_2-i-j-k  \nonumber\\
c^{n}_ i(\zeta) = \frac{ (-1)^{n+i} n! (2\zeta^2 R^2)^i}{
(2\zeta)^n (2i-n)! (n-i)!}.  
\end{eqnarray}

With these definitions we reach the main result of this work:
\begin{eqnarray}
  \label{final}
\fl
 I^{n_1 l_1 m_1}_{n_2 l_2 m_2}(\kb,\zeta_1,\zeta_2,\RI,\RJ) =  
 \sum_{{l'_1}=0}^{l_1}  (-1)^{l'_1}  i^{l_1-{l'_1}}
  \sum_{{m_1}'=-({l_1}-{l'_1})}^{{l_1}-{l'_1}} 
z^{{m_1}'}_{l_1-{l'_1}}({\bf k}) 
  \sum_{{m_1}''=-{l'_1}}^{{l'_1}} d^{{l_1} {m_1}}_{{l'_1} {m_1}'
    {m_1}''} \times \\\nonumber\fl
  \sum_{{l'_2}=0}^{l_2}  i^{l_2-{l'_2}}
  \sum_{{m_2}'=-({l_2}-{l'_2})}^{{l_2}-{l'_2}} 
z^{{m_2}'}_{l_2-{l'_2}}({\bf k}) 
  \sum_{{m_2}''=-{l'_2}}^{{l'_2}} d^{{l_2} {m_2}}_{{l'_2} {m_2}'
    {m_2}''}  
 \sum_{k=0}^{L_{<}} {\cal P}_k^{l'_1 m''_1 l'_2 m''_2} ({\bf R}) \tilde{{\cal S}}^{n_1 l_1 n_2
  l_2}_{{l'_1} {l'_2} k}({\bf k},\zeta_1,\zeta_2,\RI,\RJ), 
\end{eqnarray}
where the product of the two regular harmonics could be 
rewritten as a sum over a single harmonic, if the interest lies in the
partial-wave analysis of the Fourier transform. It can be easily
checked that (\ref{final}) reduces to the known result for the
overlap of STO in the limit of vanishing momentum $k$.  

It is now interesting to compare (\ref{final}) with the
related formula of Trivedi and Steinborn for the Fourier transform of
B-function products \cite{triv83}. At first glance the Trivedi-Steinborn
result looks more compact and involves a lower number of summations.
This is due to the favorable behavior of B-functions under the Fourier
transform. If one is interested in STO, however, as it is often the case in
quantum chemical or condensed matter problems, 
Equation (\ref{final}) provides the answer directly, while usage of the
Trivedi-Steinborn form requires a summation over several individual
integrals. Admittedly, for modest values of $n$ only a small number of
B-functions is necessary to represent a certain STO.

There is however another point which should be important in terms
of efficiency. In a
numerical quadrature a large number of function evaluations is
necessary, especially if one tries to achieve high precision.   
In the Trivedi-Steinborn form regular
spherical harmonics appear under the pending one-dimensional integral,
while the integrand in (\ref{snl}) is simpler. Moreover, the
quantity $ \tilde{{\cal S}}^{n_1 l_1 n_2
  l_2}_{{l'_1} {l'_2} k}$ does not depend on magnetic quantum
numbers and can be precomputed for every ${l'_1},{l'_2}$ and 
stored in an array of dimension $k$.

\section{Implementation details}
\label{imple}
In this paragraph we provide information on the implementation of the
derived expressions, discuss the issue of numerical stability and give
some benchmark data. 

The formulas of the last section are also valid in the special case of two
STO located on the same center, 
due to the following property of the modified McDonald function 
\begin{equation}
  \label{pmc}
  \lim_{x\to 0} x^{2\nu} \hat{k}_{-\nu}(x) = 2^{\nu-1} (\nu-1)! \quad \forall
  \nu > 0.
\end{equation}
Nevertheless, it is computationally much more efficient to replace the
STO product by a sum over single STO using Gaunt coefficients. In this
way the known analytical result for the Fourier transform of
individual STO given by Belki\'c and Taylor \cite{bel89} may be employed. Since
the routine for the computation of real Gaunt coefficients is called
extremely often also in the two-center case, an efficient strategy for
their evaluation becomes very important. We follow the recent work of Pinchon
and Hoggan \cite{pin07}, who devised a new index function to retrieve
precomputed Gaunts for complex spherical harmonics. Only those
coefficients that do not vanish due to selection rules are actually
stored initially. Real Gaunt coefficients may then be obtained as
outlined by Homeier and Steinborn \cite{hoh96}.

The remaining computational bottleneck is given by the numerical
integration. As already mentioned in the previous section, the term $
\tilde{{\cal S}}^{n_1 l_1 n_2 l_2}_{{l'_1} {l'_2} k}$ is constructed
right after looping over ${l'_1},{l'_2}$ as an one-dimensional
temporary array. The integrals over given values of $\mu,\, \nu$ and
$\alpha$ in (\ref{numu}) are computed only once and then stored,
since they appear repeatedly for different combinations of the
summation variables. For the numerical quadrature itself, we use 
adaptive integration as implemented in the {\em qag} routine of the
{\tt QUADPACK} library with a (7,15) Gauss-Kronrod rule \cite{quad}. With this approach
an accuracy of typically 14 significant figures is achieved for the basic
integrals as well as the overall Fourier transform.

 The algorithm
presented here is numerically stable for a wide range of quantum
numbers, inter-center distances and momenta ${\bf k}$. In situations
where the ratio of decay constants $\zeta_1/\zeta_2$ is large, we however do find a
significant digital erosion. For example, we found still 13 figure
accuracy for a certain integral with a decay constant ratio of 50,
which reduced to eleven figures at a ratio of 100 and finally three
figures at a ratio of 150. This drawback was also observed in related earlier
studies \cite{gro88,mon72} and possible remedies were suggested by
Homeier and Steinborn \cite{hoh92} and recently by Safouhi and Berlu
\cite{saf06}. In most real world applications the atomic
numbers of elements constituting the structure in question usually do
not differ grossly. If the interest is however in properties like
electronic excited states or polarizabilities, additional diffuse basis functions with
small decay constants are
required. In these cases a careful and more sophisticated evaluation of the basic integrals
is necessary as outlined for example by Homeier and Steinborn
\cite{hoh92}.

\begin{table}
\caption{Fourier transforms over products of  {\em normalized } STO
  which share the following parameters: $\RI$ = (0.3, -0.6, 0.9), $\RJ$
  = (1.8, 0.9, 0.1), ${\bf k}$ = (0.4, -0.7, 0.1), $\tilde{n}_1$ = 5, $\tilde{n}_2$ =
  4 (principal quantum number), $\zeta_1$ = 3.0, $\zeta_2$ = 9.0. 
\label{tab1}}
\begin{indented}
\lineup
\item[]\begin{tabular}{@{}lllllllllllll}
\br
$l_1$ & $m_1$ & $l_2$ & $m_2$ &
 \centre{9}{ $\int \dr e^{i \kb {\bf r}} \bar{\chi}_{n_1 l_1 m_1}(\rI, \zeta_1) \bar{\chi}_{n_2 l_2
    m_2}(\rJ, \zeta_2)$}  
\\
\mr
 0 & 0 & 0 & 0 & 1.3252 & 7422 & 8497 &$\times 10^{-1}$ &i& 1.8979&
 8247& 0877& 
 $\times 10^{-2}$ \\
 1 & 1 & 0 & 0 & 1.4512 & 7601 & 7773 &$\times 10^{-1}$ &i&   3.0116
 & 4031 & 2294 &   
 $\times 10^{-2}$ \\ 
1 & \-1 & 1 & \-1 & \-1.6452 & 5684 & 6177 &$\times 10^{-1}$ &\-i&
3.6525 & 2296 & 6886 & 
 $\times 10^{-2}$ \\
1 & \-1 & 1 & 0 & 3.0597 & 8029 & 2345 &$\times 10^{-2}$  &i& 7.9667
& 4828 & 0853 & 
 $\times 10^{-3}$   \\
1 & \-1 & 1 &  1 & \-6.0005 & 5763 & 7932  &$\times 10^{-2}$  &i&
1.6370 & 3748 & 7660 & 
 $\times 10^{-2}$  \\ 
2 & \-2 & 2 &  2 &  5.3441 & 5583 & 8640  &$\times 10^{-3}$  &\-i&
1.4183 & 8452 & 6288  & 
 $\times 10^{-2}$ \\
2 & \-1 &  2 &  2 & \-1.5707 & 4135 & 8199  &$\times 10^{-2}$  &i&
1.0656 & 8041 & 5381 & 
 $\times 10^{-2}$ \\
2 &  0 & 2 &  2 & \-2.3656 & 8841 & 5942 &$\times 10^{-3}$  &i&
4.7624 & 2474 & 0227& 
 $\times 10^{-3}$  \\
2 &  1 & 2 &  2 &  8.0619 & 0229 & 2047 &$\times 10^{-3}$ &i&
1.6691 & 6554 & 4101 & 
 $\times 10^{-2}$  \\
2 & 2 & 2 &  2 & \-3.2683 & 5274 & 0242 &$\times 10^{-2}$ &\-i&
9.6119 & 2920 & 0390 & 
 $\times 10^{-3}$\\
\br 
\end{tabular}
\end{indented}
\end{table}

\begin{table}
\caption{Comparison of accuracy and numerical efficiency of the
  algorithm presented in this work with the one of Trivedi and
  Steinborn in the implementation of Homeier and Steinborn
  \cite{hoh92}. Parameters for the various integrals are the same as
  in table \ref{tab1}. The provided number of significant
  digits (Digits) is the minimum of the digits for real and imaginary part,
  respectively. CPU times in ms (Time) correspond to the computation of (2$l_1$ +1)
  $\times$ (2$l_2$ +1) integrals and present an average over 1000 evaluations.  
\label{tab2}}
\begin{indented}
\lineup
\item[]\begin{tabular}{@{}llllcccc}
\br
 & & & &\centre{2}{This work} &\centre{2}{Trivedi-Steinborn} \\ 
$l_1$ & $m_1$ & $l_2$ & $m_2$ &
 Digits & Time & Digits & Time 
\\
\mr
 0 & 0 & 0 & 0    & 15 & 0.80 &14 & 0.25\\
 1 & 1 & 0 & 0    & 15 & 1.05 &14 & 0.88\\ 
1 & \-1 & 1 & \-1 & 14 & 1.62 &14 & 2.40 \\
1 & \-1 & 1 & 0   & 14 &      &13 &     \\
1 & \-1 & 1 &  1  & 14 &      &14  &        \\ 
2 & \-2 & 2 &  2  & 14 & 2.75 &13 & 10.69 \\
2 & \-1 &  2 &  2 & 13 &      &0   &    \\
2 &  0 & 2 &  2   & 13 &      &13  &    \\
2 &  1 & 2 &  2   & 13 &      &13 &   \\
2 & 2 & 2 &  2    & 13 &      &14  &\\
\br 
\end{tabular}
\end{indented}
\end{table} 
In table \ref{tab1} and \ref{tab2} we provide some benchmark results for selected
parameter values. The numerical error is estimated by a comparison
with direct three-dimensional integration (Equation (\ref{ft})) performed
with the computer algebra package {\em maple}, that features arbitrary
precision arithmetic. The CPU timings of the 
algorithm were performed on an Intel Pentium IV at 3.40GHz. The
evaluation of a Fourier transform takes roughly some hundreds of $\mu
s$ which can be compared to the computational cost of a simple overlap
integral on a similar machine, which was reported to be about three
orders of magnitude 
lower \cite{ema08}. This had to be expected, since in the latter case no
numerical quadrature is required. In addition, Equation (\ref{final}) shows
a much higher complexity than the expression for the overlap. An
important point for calculations in extended basis sets is also
apparent from table \ref{tab1}. The general computational cost increases with
increasing angular momentum, but the integrals for different
combinations of the magnetic quantum number come at little additional
cost. In fact, the CPU time per integral is {\em decreasing} with increasing
$l$. This is a consequence of the fact that the major bottleneck of
this scheme is the construction of the quantity  $\tilde{{\cal S}}^{n_1 l_1 n_2
  l_2}_{{l'_1} {l'_2} k}$ (\ref{snl}) which is independent of $m$.

In order to further explore the numerical efficiency of our approach,
we performed test calculations with the {\tt FT2B} code of Homeier,
which implements the Trivedi-Steinborn formula and is described in
detail in \cite{hoh92}. Using M\"{o}bius-transformation-based
quadrature rules, these
authors were able to handle the highly oscillatory integrand of the
remaining one-dimensional quadrature very efficiently. Utilizing the
known formulas for the conversion of B-functions to STO (see e.g.~\cite{hoh90}), we were able
to reproduce the results of table \ref{tab1}, with one
exception\footnote{The case of
  $l_1=2,\,m_1=-1,\,l_2=2,\,m_2=2$.}. The comparative timings given in
table \ref{tab2} were
performed on the same machine and with comparable code
optimization. Since the {\tt FT2B} implementation is based on complex
spherical harmonics, evaluations for different combinations of
magnetic quantum numbers were necessary to obtain Fourier transforms of real
STO. This additional effort was not included in the timings, since the
Trivedi-Steinborn formula might be equally well formulated in real
spherical harmonics. 

We find for the special choice of quantum numbers given in table
\ref{tab2}, that the {\tt FT2B} implementation is superior to
our approach for individual integrals by roughly a factor of
four. In general, one STO product may be represented by $\left(\lfloor
(\tilde{n}-l)/2 \rfloor +1\right)^2$ B-function products, so that this
result is strongly parameter dependent. In applications one is usually
interested in the full set of integrals for different combinations of
$m$-values and here our approach is numerically more efficient as
table \ref{tab2} shows. These computational savings will moreover
increase with increasing angular momentum.

Code improvements are possible for both the B-function approach as
well as for our scheme. Homeier mentions in his dissertation
\cite{hoh90}, that storage of some intermediate quantities might
improve the performance for higher angular momentum. Our
implementation might benefit from the M\"{o}bius quadrature put forward in
\cite{hoh92}. While the integrand is evaluated at 36 points in the
{\tt FT2B} implementation, our adaptive integration requires 135
points for the same precision. A speed-up of a factor of four seems
therefore achievable.

\section{Summary}
In this work a compact general purpose formula for the Fourier transform of
STO products with arbitrary quantum numbers and geometrical parameters
was derived. We highlighted the relation to earlier work based on
B-functions and found differences that are relevant for the numerical
efficiency. It should be stressed that the derivation presented here
is completely independent. Moreover, the final formula can not be
reduced to the Trivedi-Steinborn result by a mere transformation from
B-functions to STO. Regarding numerical stability which is often an
issue in STO related studies \cite{bar02}, we achieved in general a
completely satisfying accuracy apart from the known problems with very
unsymmetric orbital products.  We expect that the typical computational
cost of several $\mu$s per integral allows for a very efficient evaluation of the
notoriously complicated four-center electron repulsion integrals. The
Fourier transform technique hence provides a viable alternative 
to existing direct methods in the field.

\ack
We would like to thank Dr.~Homeier for helpful discussions and also
for providing us with a copy of his
{\tt FT2B} code.

\appendix

\section{Some derivatives and further definitions}
\label{further}
The derivative of the modified McDonald function $\hat{k}_\nu(x)$ has
the following simple form 
\begin{equation}
  \label{dmc}
  \frac{d \hat{k}_\nu(x)}{d x} = - x \hat{k}_{\nu-1}(x).
\end{equation}
In order to evaluate the quantity $ h^{n_1 l_1}_{n_2 l_2}$ in
(\ref{hterm}) an expression for the repeated action of the operator
$-\frac{1}{\zeta} \frac{\partial}{\partial \zeta}$ on $\hat{k}$ is
required. Straightforward differentiation leads to
\begin{eqnarray}
\fl
 \left(-\frac{1}{\zeta_1} \frac{\partial}{\partial
     \zeta_1}\right)^{l}  \hat{k}_{\nu}\left(R \sqrt{\zeta_u^2 + k^2
    u (1-u)} \right) &= R^{2 l} u^l  \hat{k}_{\nu-l}\left(R \sqrt{\zeta_u^2 + k^2
    u (1-u)} \right). \nonumber\\
\fl 
 \left(-\frac{1}{\zeta_2} \frac{\partial}{\partial
     \zeta_2}\right)^{l}  \hat{k}_{\nu}\left(R \sqrt{\zeta_u^2 + k^2
    u (1-u)} \right) &= R^{2 l} (1-u)^l  \hat{k}_{\nu-l}\left(R \sqrt{\zeta_u^2 + k^2
    u (1-u)} \right).   \label{a1}
\end{eqnarray}
The action of the operator $-\frac{\partial}{\partial \zeta}$ is more
involved but can be reduced to (\ref{a1}).
\begin{eqnarray}
  \label{a2}
  \left(-\frac{\partial}{\partial \zeta}\right)^n =
  \frac{n!(-1)^n}{(2\zeta)^n} \sum_{i=\lfloor\frac{n+1}{2}\rfloor}^{n} 
 \frac{(-2\zeta^2)^i}{(2i-n)! (n-i)!}  \left(-\frac{1}{\zeta}\frac{\partial}{\partial \zeta}\right)^i, 
\end{eqnarray}
which gives rise to the definitions of the coefficients $c^{n}_
i(\zeta)$ in (\ref{numu}).

We now prove (\ref{a2}) by using induction. The induction basis for
$n=1$ is trivial. We further have ($\delta_{i,k}$ denoting the
Kronecker delta)
\begin{eqnarray}
  \label{ap1}
\fl
\left(-\frac{\partial}{\partial \zeta}\right)
\left(-\frac{\partial}{\partial \zeta}\right)^n &= \sum_{i=\lfloor
  \frac{n+1}{2} \rfloor}^n \frac{n! (-1)^{n+1+i} 2^{i-n} }{ (2i-n-1)!
  (n-i)!} \zeta^{2i-n-1}
\left(1-\delta_{i,n/2}\right)\left(-\frac{1}{\zeta}\frac{\partial}{\partial
    \zeta}\right)^i  \nonumber\\
&+  \sum_{i'=\lfloor
  \frac{n+1}{2} \rfloor + 1}^{n+1} \frac{n! (-1)^{n+1+i'} 2^{i'-n-1} }{ (2i'-n-2)!
  (n+1-i')!} \zeta^{2i'-n-1}
\left(-\frac{1}{\zeta}\frac{\partial}{\partial
    \zeta}\right)^{i'},
\end{eqnarray}
where we used the induction hypothesis and changed the summation index
to $i'=i+1$ in the second sum. Separating the term for the lower limit $i=\lfloor
  \frac{n+1}{2} \rfloor$ in the first sum and the upper limit $i'=n+1$
  in the second sum, we arrive at
  \begin{eqnarray}
    \label{ap2}
    \fl
\left(-\frac{\partial}{\partial \zeta}\right)^{n+1} &= 
 \frac{n! (-1)^{n+\lfloor
  \frac{n+1}{2} \rfloor + 1} 2^{\lfloor
  \frac{n+1}{2} \rfloor-n} }{ (2\lfloor
  \frac{n+1}{2} \rfloor-n-1)!
  (n-\lfloor
  \frac{n+1}{2} \rfloor)!} \zeta^{2\lfloor
  \frac{n+1}{2} \rfloor-n-1}
\left(1-\delta_{\lfloor
  \frac{n+1}{2} \rfloor,n/2}\right)\left(-\frac{1}{\zeta}\frac{\partial}{\partial
    \zeta}\right)^{\lfloor
  \frac{n+1}{2} \rfloor} \nonumber\\
&+ \sum_{i=\lfloor
  \frac{n+1}{2} \rfloor+1}^n \frac{(n+1)! (-1)^{n+1+i} 2^{i-n-1} }{ (2i-n-1)!
  (n+1-i)!} \zeta^{2i-n-1}
\left(-\frac{1}{\zeta}\frac{\partial}{\partial
    \zeta}\right)^i  \nonumber\\
&+ \zeta^{n+1} \left(-\frac{1}{\zeta}\frac{\partial}{\partial
    \zeta}\right)^{n+1}. 
  \end{eqnarray}
For even $n$, we have $\lfloor \frac{n+1}{2} \rfloor = n/2$, and the
first term in (\ref{ap2}) vanishes. Since in this case $\lfloor \frac{n+1}{2}
\rfloor +1 = \lfloor \frac{n+2}{2} \rfloor$, it follows: 
\begin{equation}
  \label{ap3}
\fl
\left(-\frac{\partial}{\partial \zeta}\right)^{n+1} =   \sum_{i=\lfloor
  \frac{n+2}{2} \rfloor}^{n+1} \frac{(n+1)! (-1)^{n+1+i} 2^{i-n-1} }{ (2i-n-1)!
  (n+1-i)!} \zeta^{2i-n-1}
\left(-\frac{1}{\zeta}\frac{\partial}{\partial
    \zeta}\right)^i.   
\end{equation}
For odd $n$, we have $\lfloor \frac{n+1}{2} \rfloor = (n+1)/2$, as well
as $\lfloor \frac{n+1}{2}
\rfloor +1 = \lfloor \frac{n+2}{2}\rfloor + 1$. If we now extend the
second sum in (\ref{ap2}) to the lower limit $ i= \lfloor
\frac{n+2}{2}\rfloor$ the compensating term cancels exactly the first
term in  (\ref{ap2}). Also in this case we therefore arrive at
(\ref{ap3}), that is the hypothesis for $n+1$, which was to be demonstrated.

Derivatives with respect to the inter-center distance $R$ are likewise
readily obtained:
\begin{equation}
  \label{a3}
\fl
 \left(\frac{1}{R} \frac{\partial}{\partial
     R}\right)^{l} R^{-2\nu} \hat{k}_{\nu}\left(R \sqrt{\zeta_u^2 + k^2
    u (1-u)} \right) = (-1)^l R^{-2(\nu + l)} \hat{k}_{\nu + l}\left(R \sqrt{\zeta_u^2 + k^2
    u (1-u)} \right),  
\end{equation}
where we have used the following recursion:
\begin{equation}
  \label{a4}
\hat{k}_{\nu+1}(x) = 2\nu \hat{k}_{\nu}(x) + x^2  \hat{k}_{\nu - 1}(x).    
\end{equation}

A combination of (\ref{a1}) to (\ref{a3}) leads to second line in
(\ref{snl}).

Finally, we define the coefficients $c^{l_1 m_1 l_2 m_2}_{l_3 m_3}$
that appear in (\ref{pl}). A product of two regular harmonics with
same argument can
be linearized as follows:
\begin{equation}
  \label{zpr}
  z^{m_1}_{l_1} ({\bf R})  z^{m_2}_{l_2} ({\bf R}) = \sum_{l_3}
  \sum_{m_3} c^{l_1 m_1 l_2 m_2}_{l_3 m_3} z^{m_3}_{l_3} ({\bf R})
  R^{l_1 +l_2 -l_3}.
\end{equation}
In terms of the Gaunt-like coefficients for the real {\em unnormalized}
spherical harmonics of (\ref{zlm}) 
\begin{equation}
  \label{gau}
  \left[ l_1 m_1 | l_2 m_2 | l_3 m_3 \right] = \int
  \tilde{Y}_{l_1 m_1}(\Omega) \tilde{Y}_{l_2 m_2}(\Omega) \tilde{Y}_{l_3 m_3}(\Omega) d\Omega,
\end{equation}
these coefficients read 
\begin{equation}
  \label{fin}
   c^{l_1 m_1 l_2 m_2}_{l_3 m_3} = \left(\frac{ (2l_3 + 1)}{2\pi
       (1+\delta_{m_3 0})}
    \frac{(l_3 -|m_3|)!}{ (l_3 +|m_3|)!}\right)  \left[ l_1 m_1 | l_2 m_2 | l_3 m_3 \right].
\end{equation}

Please note that notation (\ref{gau}) differs from the one usually
employed for Gaunt coefficients  \cite{gau29}. The linearization formula (\ref{zpr})
may be considerably simplified by taking advantage of the selection
rules for the Gaunt-like coefficients (\ref{gau}), which were
discussed by Homeier and Steinborn \cite{hoh96}: 
\begin{eqnarray}
  \label{aaa}
   z^{{m_1}}_{{l_1}}({\bf R})
  z^{{m_2}}_{{l_2}}({\bf R}) = \sum_{l=0}^{L_{<}} \sum_m c^{l_1 m_1
    l_2 m_2}_{l_1+l_2-2l m} z^m_{l_1 + l_2 - 2l} ({\bf R}) R^{2l}\nonumber\\
  m\in \left\{ m_1+m_2, m_1-m_2, -m_1+m_2, -m_1-m_2\right\} 
\end{eqnarray}

\section{Leibniz theorem for regular harmonics}
\label{leib}
Here we prove Eq. (\ref{dzlm}) of section \ref{trahigh}
\begin{equation}
  \label{app1}
   z^{m}_{l}(\nabla) (f\, g) = \sum_{l'=0}^l 
  \sum_{m'=-(l-l')}^{l-l'} 
  \sum_{m''=-l'}^{l'} d^{l m}_{l' m' m''} \left(z^{m'}_{l-l'}(\nabla)\, f \right) 
  \left(z^{m''}_{l'}(\nabla)\, g \right).    
\end{equation}

The regular harmonic $z^{m}_{l}(\nabla)$ is given in cartesian form as
\begin{equation}
  \label{apa2}
  z^{m}_{l}(\nabla) = \sum_{i=0}^l \sum_{j=0}^{l-i} C^{lm}_{i j}
  \left(\frac{\partial}{\partial x}\right)^{(i)} 
  \left(\frac{\partial}{\partial y}\right)^{(j)} 
  \left(\frac{\partial}{\partial z}\right)^{(l-i-j)}, 
\end{equation}
where $\left( \partial/\partial x\right)^{(n)}$ denotes the $n$-th
partial derivative with respect to $x$ and the coefficients $C^{lm}_{i
  j}$ are known constants (see e.g.~\cite{fer01}, Eqs. 3 and
4). Applying the Leibniz theorem for the differentiation of products
we have
\begin{eqnarray}
  \label{apa3}
 \fl  z^{m}_{l}(\nabla) (f\, g) = \sum_{i=0}^l \sum_{j=0}^{l-i} C^{lm}_{i
     j} \sum_{i'=0}^i \sum_{j'=0}^j \sum_{k'=0}^{l-i-j}
   {i \choose i'}
   {j \choose j'}
   {l-i-j \choose k'} \times\\\nonumber
\left[ 
  \left(\frac{\partial}{\partial x}\right)^{(i-i')} 
  \left(\frac{\partial}{\partial y}\right)^{(j-j')} 
  \left(\frac{\partial}{\partial z}\right)^{(l-i-j-k')}
f
\right] 
\left[
  \left(\frac{\partial}{\partial x}\right)^{(i')} 
  \left(\frac{\partial}{\partial y}\right)^{(j')} 
  \left(\frac{\partial}{\partial z}\right)^{(k')}
g
 \right]. 
\end{eqnarray}
Using now the completeness of the regular harmonics we may expand the
product $x^i y^j z^k$ into harmonics of angular momentum $l=i+j+k$
\begin{eqnarray}
  \label{apa4}
  x^i y^j z^k = \sum_{m=-l}^{l} B^l_m
  z^{m}_{l}({\bf r}) \quad ; \quad l=i+j+k \\\nonumber
 B^l_m = \int x^i y^j z^k r^{-2l}  z^{m}_{l}({\bf r})\, d\Omega.
\end{eqnarray}
Inserting this expansion into (\ref{apa3}) we find
\begin{eqnarray}
  \label{apa5}
 \fl  z^{m}_{l}(\nabla) (f\, g) = \sum_{i=0}^l \sum_{j=0}^{l-i} C^{lm}_{i
     j} \sum_{i'=0}^i \sum_{j'=0}^j \sum_{k'=0}^{l-i-j}
   {i \choose i'}
   {j \choose j'}
   {l-i-j \choose k'} \times\\\nonumber
\left[ 
\sum_{m'=-(l-i'-j'-k')}^{l-i'-j'-k'} B^{l-i'-j'-k'}_{m'}  z^{m'}_{l-i'-j'-k'}(\nabla)
\,f
\right]
\left[
\sum_{m''=-(i'+j'+k')}^{i'+j'+k'} B^{i'+j'+k'}_{m''}  z^{m''}_{i'+j'+k'}(\nabla)
\,g
 \right],
\end{eqnarray}
which can be simplified after changing the summation order according
to 
\begin{equation}
  \label{apa6}
  \sum_{a=0}^{a'} \sum_{b=0}^{b'}  \sum_{c=0}^{c'} F(a,b,c) =
  \sum_{a=0}^{a'+b'+c'}  \sum_{b=0}^{{\rm min}(a,b')} 
    \sum_{c={\rm max}(0,a-b-a')}^{{\rm min}(a-b,c')} F(a-b-c,b,c),
\end{equation}
for arbitrary $F$. With the help of the coefficients $A^{l'}_{lm}$
\begin{equation}
  \label{apa7}
A^{l'}_{lm}= \sum_{i=0}^l \sum_{j=0}^{l-i} C^{lm}_{i
     j} \sum_{j'=0}^{{\rm min}(l',j)} \sum_{k'={\rm max}(0,l'-j'-i')}^{{\rm min}(l'-j',l-i-j)}
   {i \choose l'-j'-k'}
   {j \choose j'}
   {l-i-j \choose k'}, 
\end{equation}
with $l'=i'+j'+k'$, we finally arrive at 
\begin{equation}
  \label{apa8}
  z^{m}_{l}(\nabla) (f\, g)  = \sum_{l'=0}^l  A^{l'}_{lm} \sum_{m'=-(l-l')}^{l-l'}
  \sum_{m''=-l'}^{l'} \left[ 
B^{l-l'}_{m'}  z^{m'}_{l-l'}(\nabla)
\,f
\right]
\left[
 B^{l'}_{m''}  z^{m''}_{l'}(\nabla)
\,g
 \right],
\end{equation}
which is equivalent to (\ref{dzlm}) if we set $d^{l m}_{l' m' m''} = A^{l'}_{lm}
B^{l'}_{m''} B^{l-l'}_{m'}$.

\section{Proof of equation (\ref{hobson})}
\label{hobs} 
An old theorem given by Hobson \cite{hob65} (see also \cite{wen83})
states that if $R=(X^2+Y^2+Z^2)^{\hal}$ and $H(X,Y,Z)$ is a
homogeneous polynomial of degree $l$ in the $X,Y,Z$ variables, then:
\begin{equation}
  \label{aph1}
  H\left( \frac{\partial}{\partial X},\frac{\partial}{\partial
      Y},\frac{\partial}{\partial Z}\right) f(R) = \sum_{k=0}^l
  \frac{2^{l-2k}}{k!} \left[ \nabla^{2k} H(X,Y,Z)\right]
  \left( \frac{\partial}{\partial R^2}\right)^{l-k} f(R).   
\end{equation}
Taking into account that $\frac{\partial}{\partial X} =
-\frac{\partial}{\partial X_I} = \frac{\partial}{\partial X_J}$ from $
{\bf R} = {\bf R}_J -  {\bf R}_I$, we may apply (\ref{aph1}) to the
product of the regular harmonics $ z^{{m_1}}_{{l_1}}(\nabla)$
and $z^{{m_2}}_{{l_2}}({\nabla})$, which is a homogeneous polynomial of
degree $l_1 + l_2$. Hence,
\begin{eqnarray}
\fl
\nonumber
   z^{{m_1}}_{{l_1}}(\nabla_I)
  z^{{m_2}}_{{l_2}}({\boldmath \nabla}_J) f(R) &= 
(-1)^{l_1}
  \sum_{k=0}^{l_1+l_2} \frac{2^{l_1+l_2-2k}}{k!} \left[ \nabla^{2k}
    z^{{m_1}}_{{l_1}}({\bf R})
  z^{{m_2}}_{{l_2}}({\bf R}) \right]
\left(\frac{\partial}{\partial R^2}\right)^{l_1 +l_2 -k}\, f(R)
\\
  \label{aph2}
&= (-1)^{l_1}
  \sum_{k=0}^{l_1+l_2} \frac{2^{-k}}{k!} \left[ \nabla^{2k}
    z^{{m_1}}_{{l_1}}({\bf R})
  z^{{m_2}}_{{l_2}}({\bf R}) \right]
\left(\frac{1}{R} \frac{\partial}{\partial R}\right)^{l_1 +l_2 -k}\,
f(R),
\end{eqnarray}
which is the same as (\ref{hobson}) apart from the upper 
limit in the sum over $k$.
In Ref.~\cite{ema08} it was shown that 
\begin{equation}
  \label{aph3}
 \nabla^{2k} z^m_p({\bf R}) R^{2l} =
\left\{ 
\begin{array}{c@{\quad :\quad}l}
 \frac{2^{2k} l! \Gamma(p+l+3/2)
  }{(l-k)! \Gamma( p+l-k+3/2)}  z^m_p({\bf R}) R^{2l-2k} & k \le l \\
 0 & k > l 
\end{array}
\right.
\end{equation}
Combining (\ref{aph2}) with (\ref{aaa}) and  (\ref{aph3}) we arrive
at (\ref{hobson}), (\ref{hobson2}) and (\ref{pl}) of the main paper.


\vspace*{2cm}
 

\begin{thebibliography}{99}
\bibitem{bon64} R A   Bonham, J L  Peacher and H L  Cox 1964 {\it J. 
  Chem.  Phys.}  {\bf 40} 3083 
\bibitem{bon65} R A  Bonham 1965 {\it J.  Phys.  Soc.  Jpn.}  {\bf 20} 2260
\bibitem{hed65} In Hedin's GW approximation, the self-energy operator of
  quasiparticle theory is given by the product of the one-particle Greens
  function G and the screened Coulomb operator W. 
 See L  Hedin 1965 {\it Phys.  Rev.}  A {\bf 139} 796 
\bibitem{roh98}   M  Rohlfing and S G Louie 1998 {\it Phys.  Rev.
    Lett.}  {\bf 80} 3320  
\bibitem{roh00}   M Rohlfing 2000 {\it Int.  J.  Quant.  Chem.}  {\bf 80} 807
\bibitem{oht01}   Y  Ohta, J Maki, T Yoshimoto, Y Shigeta, H Nagao and
  K Nishikawa 2001 {\it  Int.  J.  Quant.  Chem.}  {\bf 84} 348 
\bibitem{nie05} T A Niehaus, M Rohlfing, F Della Sala, A Di Carlo,
  and T Frauenheim 2005 {\it Phys. Rev.} A {\bf 71} 022508 
\bibitem{nea06} J B  Neaton, M S  Hybertsen and S G Louie
  2006 {\it Phys.  Rev.  Lett.}  {\bf 97} 216405 
\bibitem{pop70}  J A  Pople and J L  Beveridge 1970 {\it Approximate
    Molecular Orbital 
    Theory } (New York: McGraw Hill)
\bibitem{pis80} C Pisani and R Dovesi  1980 {\it Int.  J.  Quant.  Chem.} 
  {\bf 17} 501
\bibitem{ben73}  J Bentley and R F Stewart 1973 {\it J. Comput. Phys.}
  {\bf 11} 127
\bibitem{jun80} B R Junker 1980 {\em J. Phys. B} {\bf 13} 1049
\bibitem{stra87} J C Straton 1987 {\em Phys. Rev. A} {\bf  35} 2729
\bibitem{stra88} J C Straton 1988 {\em Phys. Rev. A} {\bf  37} 4531
\bibitem{mon76} H J  Monkhorst and J D  Pack 1976 {\it Phys.  Rev.}  B
  {\bf 13} 5188
\bibitem{triv83} H P  Trivedi and E O  Steinborn 1983 {\it Phys.  Rev.}  A
  {\bf 27} 670 
\bibitem{gro88} J  Grotendorst and E O  Steinborn 1988 {\it Phys.  Rev.}  A
  {\bf 38} 3857
\bibitem{gro85} J Grotendorst and E O Steinborn 1985 {\em J. Comput.
    Phys.} {\bf 61} 195
\bibitem{hoh90} H H H Homeier 1990 {\em
    Integraltransformationsmethoden und Quadraturverfahren f\"{u}r
    Molek\"{u}lintegrale mit B-Funktionen}  (S. Roderer: Regensburg)
\bibitem{ema08} I  Ema, R  L\'opez, J J  Fern\'andez, 
  G Ram\'irez  and  J F Rico 2008 {\it Int.  J.  Quant.  Chem.} {\bf 108} 25 
\bibitem{vel01} G  te Velde, F M  Bickelhaupt, E J  Baerends, C 
  Fonseca Guerra, S J A  van Gisbergen, J G  Snijders and  T 
  Ziegler 2001 {\it J.  Comp.  Chem.}  {\bf 22} 931 
\bibitem{fer89} G W  Fernando, J W  Davenport, R E  Watson and
  M Weinert 1989 {\it Phys. Rev.} B {\bf 40} 2757
\bibitem{gra80}  I S Gradshteyn and I M Ryzhik 1980 {\it Table of
    Integrals, Series and Products} (London: Academic Press) Eq. 8.812
\bibitem{bar51} M P Barnett and C A Coulson 1951 {\em
    Phil. Trans. Roy. Soc. London A} {\bf  243} 221
\bibitem{gui79} G  Guidotti, G P  Arrighini and F  Marinelli 1979
 {\it Theor.  Chim.  Acta.}  {\bf 53} 165  
\bibitem{fie80} G Fieck 1980 {\em Theor. Chim. Acta} {\bf 54} 323 
\bibitem{wen83} E J Weniger and E O Steinborn 1983 {\em J. Chem. Phys.}
  {\bf 78} 6121  
\bibitem{fer00} J F Rico, J J  Fern\'andez, R  L\'opez,
  and G Ram\'irez 2000 {\it Int.  J.  Quant.  Chem.} {\bf 78} 83 
\bibitem{fer01}  J F Rico,  R L\'opez,
  and G Ram\'irez 2001 {\it J.  Mol.  Struct.  THEOCHEM} {\bf 537} 27
\bibitem{wen05}  E J Weniger 2005 {\em  Collect. Czech. Chem. Commun.}
  {\bf 70} 1225
\bibitem{ric07} J F Rico, I Ema, R L\'opez, G Ram\'irez
  and K Ishida  2008 {\it Recent Advances in Computational Chemistry,
    Molecular Integrals over Slater Orbitals} ed T Ozdogan
  and B Ruiz (Kerala, India: Transworld Research Network)
 chap. 5
\bibitem{gra80eq} I S Gradshteyn and I M Ryzhik 1980 {\it Table of
    Integrals, Series and Products} (London: Academic Press)
  Eq. 6.726.3
\bibitem{dun90} B I Dunlap 1990 {\em Phys. Rev. A} {\bf 42} 1127 
\bibitem{hob65} E W Hobson 1965 {\it The Theory of Spherical and
    Ellipsoidal Harmonics} (New York: Chelsea) p. 127, Eq. 7
\bibitem{bel89} D Belki\'c and H S Taylor 1989 {\it Physica Scripta} {\bf
    39} 226
\bibitem{pin07} D Pinchon and P E  Hoggan 2007 {\it Int.  J.  Quant.  Chem. }
  {\bf 107} 2186 
\bibitem{hoh96} H H H  Homeier and E O  Steinborn 1996 {\it J.  Mol. 
  Struct.  Theochem} {\bf 368} 31 
\bibitem{quad} R Piessens, E De Doncker-Kapenga and
  C W \"Uberhuber 1983 {\it  QUADPACK: a subroutine package for automatic
  integration} (Berlin, Heidelberg, New York, Tokyo: Springer)

\bibitem{mon72} H J  Monkhorst and F E  Harris 1972 {\it Int.  J.  Quant.  Chem.} 
  {\bf 6} 601
\bibitem{hoh92} H H H Homeier and E O Steinborn. 1992 {\em
    Int. J. Quantum Chem.} {\bf 41} 399
\bibitem{saf06}  H. Safouhi and L. Berlu 2006 {\em J. Comput. Phys.}
  {\bf 216} 19
\bibitem{bar02} M B   Barnett 2002 {\it Theor.  Chem.  Acc.}  {\bf
    107} 241 
\bibitem{gau29} J A Gaunt 1929 {\em Phil. Trans. Roy. Soc. A}
  {\bf 228} 151
\end{thebibliography}
\end{document}